\documentclass{aa}
\usepackage{graphicx}
\usepackage{comment}
\usepackage{txfonts}
\usepackage{xspace}
\usepackage{ulem}
\usepackage{color}
\usepackage{hyperref}
\usepackage[switch]{lineno}
\usepackage{orcidlink}
\usepackage{linenoaa} 
\newcommand {\be}{\begin {equation}}
\newcommand {\ee}{\end {equation}}

\newcommand{\bea}{\begin{eqnarray}}
\newcommand{\eea}{\end{eqnarray}}

\newcommand{\Rns}{R_{\rm NS}}

\newcommand {\flux}{erg~s$^{-1}$~cm$^{-2}$\xspace}
\newcommand {\lum}{erg~s$^{-1}$\xspace}

\newcommand{\freddi}{\textsc{freddi}\xspace}
\newcommand {\nustar}{\textit{NuSTAR}\xspace}

\newcommand {\gro}{\mbox{GRO~J1008$-$57}\xspace}
\newcommand {\src}{\mbox{4U~0115+63}\xspace}
\newcommand {\srcV}{\mbox{V~0332$+$53}\xspace}
\newcommand {\smc}{\mbox{SMC~X-2}\xspace}
\newcommand {\sw}{\mbox{Swift~J0243$+$6124}\xspace}

\begin{document} 

\title{Real-time observations of the transition to the quiescent state in an accreting magnetised neutron star: No propeller required?}

\author{%\small 
Sergey~S.~Tsygankov\inst{\ref{in:UTU}, \ref{in:Tub},\ref{in:IHEP}}\orcidlink{0000-0002-9679-0793}
\and Galina Lipunova\inst{\ref{in:FAU},\ref{in:MPIfR}}
\and Valery~F.~Suleimanov\inst{\ref{in:Tub}}\orcidlink{0000-0003-3733-7267}
\and Alexander~Salganik\inst{\ref{in:UTU}}\orcidlink{0000-0003-2609-8838} 
\and Alexander~A.~Mushtukov\inst{\ref{in:UCL},\ref{in:Oxford}}\orcidlink{0000-0003-2306-419X}
\and Sofia~V.~Forsblom\inst{\ref{in:UTU}}\orcidlink{0000-0001-9167-2790}
\and Andrey~S.~Tavleev\inst{\ref{in:Tub}}\orcidlink{0000-0001-6842-7383}
\and Aleksei~V.~Kuzin\inst{\ref{in:Tub}}
\and Juri~Poutanen \inst{\ref{in:UTU}}\orcidlink{0000-0002-0983-0049}
}
        
\institute{Department of Physics and Astronomy, FI-20014 University of Turku,  Finland \label{in:UTU} \\ \email{sergey.tsygankov@utu.fi}
\and
Institut f\"ur Astronomie und Astrophysik, Universit\"at T\"ubingen, Sand 1, D-72076 T\"ubingen, Germany \label{in:Tub} 
\and 
Key Laboratory of Particle Astrophysics, Institute of High Energy Physics, Chinese Academy of Sciences, Beijing 100049, China\label{in:IHEP} 
\and
Dr. Karl Remeis-Observatory and Erlangen Centre for Astroparticle Physics, Friedrich-Alexander Universität Erlangen-Nürnberg,
Sternwartstr. 7, 96049 Bamberg, Germany
\label{in:FAU}
\and
Max Planck Institut f\"ur Radioastronomie, Auf dem H\"ugel 69, D-53121 Bonn, Germany \label{in:MPIfR}
\and
Mullard Space Science Laboratory, University College London, Holmbury St. Mary, Surrey RH5 6NT, UK  \label{in:UCL}
\and 
Astrophysics, Department of Physics, University of Oxford, Denys Wilkinson Building, Keble Road, Oxford OX1 3RH, UK \label{in:Oxford}
}
          
\titlerunning{Disc instability in action in XRPs}
\authorrunning{S.~S.~Tsygankov et al.}

\date{25.07.2026}

\abstract{The final stages of outbursts in transient X-ray pulsars (XRPs), which are characterised by a significant decline in the mass accretion rate, provide valuable insight into the physics of the accretion disc and its interaction with the strong magnetic field of the neutron star (NS). In particular, the `propeller effect', or centrifugal inhibition of accretion, has been proposed as a key mechanism governing both the onset luminosity and the timescale of the rapid transition to the quiescent state. In addition, it offers an independent method for estimating the magnetic field strength of the NS. On the other hand, the decrease in the mass accretion rate itself is driven by processes occurring in the accretion flow at larger distances from the NS. Recovering the information encoded in the light curve therefore requires sensitive high-cadence X-ray monitoring capable of capturing the rapid and often unpredictable transition from the accreting regime to the quiescent regime. 
In this study, we present the results of the first comprehensive monitoring campaign that tracks the entire transition to  quiescence in the transient XRP \src utilising observations by the NICER X-ray telescope. 
We show that the observed behaviour can be explained by the thermal-viscous disc instability model (DIM), with the emission observed immediately after an outburst possibly arising from the ongoing accretion from the recombined (`cold') disc and the subsequent quiescent emission being produced by the cooling NS.  
We further applied this model to a larger sample of XRPs encompassing a broad range of physical parameters. 
Ultimately, our findings indicate that the temporal behaviour of XRPs, including the quiescent state, can be consistently explained within the DIM framework without requiring the propeller effect as the primary mechanism governing the observed transition.
}

\keywords{accretion, accretion discs -- magnetic fields -- pulsars: individual: 4U 0115+63 -- stars: neutron -- X-rays: binaries}

\maketitle

\nolinenumbers
\modulolinenumbers[5]

%
%-------------------------------------------------------------------
%========================================
%%%%%%%%%%%%%%%%%%%%%%%%%%%%%%%%%%%%%%%%%%%%%%%%%%%%%%%%%%%%%%%%%%%%%%%%%%%%%%
%% Introduction                                                             %%
%%%%%%%%%%%%%%%%%%%%%%%%%%%%%%%%%%%%%%%%%%%%%%%%%%%%%%%%%%%%%%%%%%%%%%%%%%%%%%
\section{Introduction} 
\label{sec:intro}

The variability of the observed flux from accreting compact objects on all timescales encodes a plethora of information on the physical conditions in the accretion disc and processes of radiation generation and propagation in the immediate vicinity of the central object. In addition, accretion onto highly magnetised neutron stars (NSs), X-ray pulsars \citep[XRPs; see][for a review]{Mushtukov_Tsygankov_2024_Review}, allows us to investigate poorly known physics of the interaction of the astrophysical plasma with a very strong magnetic field. 
Such studies became possible just recently thanks to the availability of sensitive and flexible X-ray telescopes able to monitor transient XRPs throughout a wide luminosity range. 

One of the most straightforward effects expected from the interaction between the accretion disc and the NS magnetosphere is the so-called `propeller effect' \citep{1975A&A....39..185I}. This effect arises due to the centrifugal inhibition of accretion in transient XRPs at low mass accretion rates, when the magnetospheric radius exceeds the corotation radius, defined as the distance at which the Keplerian angular velocity of the accretion flow equals the spin angular velocity of the NS.
In this situation, one may expect a sharp drop in the observed flux from the source.
Observational evidence of the propeller effect in magnetised NSs has been reported for several objects, manifested as a sudden drop in flux observed in various types of systems, including the accreting millisecond pulsar SAX~J1808.4$-$3658 \citep{2008ApJ...684L..99C}, bursting X-ray pulsar GRO~J1744$-$28 \citep{1997ApJ...482L.163C}, several classical XRPs (\src, \citealt{2001ApJ...561..924C,2016A&A...593A..16T}; \srcV, \citealt{1986ApJ...308..669S,2016A&A...593A..16T}; SMC~X-2, \citealt{2017ApJ...834..209L}), and even the pulsating ultra-luminous X-ray source M82~X-2 \citep{2016MNRAS.457.1101T}. Additionally, \cite{2018A&A...610A..46C} also considered a knee in the light curves of several classes of objects (low-mass X-ray binaries, high-mass X-ray binaries, intermediate polars, and young stellar objects)  as a manifestation of the propeller effect.

However, the propeller effect is not the only factor that can potentially affect the light curves of accreting objects. 
A gradual decline in brightness on timescales of several tens of days is a generic characteristic of the depletion of a mass reservoir surrounding a stellar-mass central object on viscous timescales. 
The physical conditions within the accretion disc itself, particularly the ionisation state of the plasma, play a crucial role in shaping the observed light curve. 
This is a basic component of the disc instability model (DIM), originally suggested for accreting cataclysmic variables  \citep{Hoshi1979,MeyerMH1981,Smak1982}. 
Here, significant changes in opacity and viscosity due to changes in the ionisation state of the disc plasma are linked to variations in the accretion rate \citep[see e.g.][for a review]{2001NewAR..45..449L}. 
Particularly, the temperature of the hot, ionised accretion disc is expected to inevitably decrease at the declining phases of an outburst. 
At some point, it may fall below a critical value of the order of $10^4$~K, triggering a transition to a cooler, partially neutral state. This transition leads to a sharp drop in viscosity and the formation of a cooling front that propagates through the disc, causing a rapid decrease in the mass accretion rate and, consequently, in the observed luminosity.

The evolution of the accretion rate during the disappearance of the hot zone in a finite amount of time corresponds to the characteristic steepening of the light curve plotted in log-linear coordinates \citep{1998MNRAS.293L..42K}.
Furthermore, the speed of the cooling front propagation can increase when the irradiation-controlled stage ends. 
This is connected to a further steepening of the light curve.
For instance, it was shown that the rapid decline in flux observed during the final stages of outbursts from the accreting non-magnetised NS Aql~X-1 can be attributed solely to the radial shrinkage of the hot (fully ionised) portion of the accretion disc  \citep{Lipunova+2022,2024ApJ...961..252C}. 
Furthermore, in the context of transient XRPs, the DIM has been shown to determine the mass accretion rate in the quiescent state in the long-period sources where recombination of the whole disc is expected to occur before a possible transition to the propeller regime \citep{2017A&A...608A..17T}. 
It is worth noting that a fast flux drop in the DIM requires significant cooling of the accretion disc below the critical temperature, and therefore, it is only possible at sufficiently low mass accretion rates. 
It can thus hardly explain propeller-like variability in ultra-luminous X-ray sources \citep{2016MNRAS.457.1101T}.

%%%%%%%%%%%%%%%%%%%%%%%%%%
\begin{figure}
\centering
\includegraphics[width=0.98\linewidth]{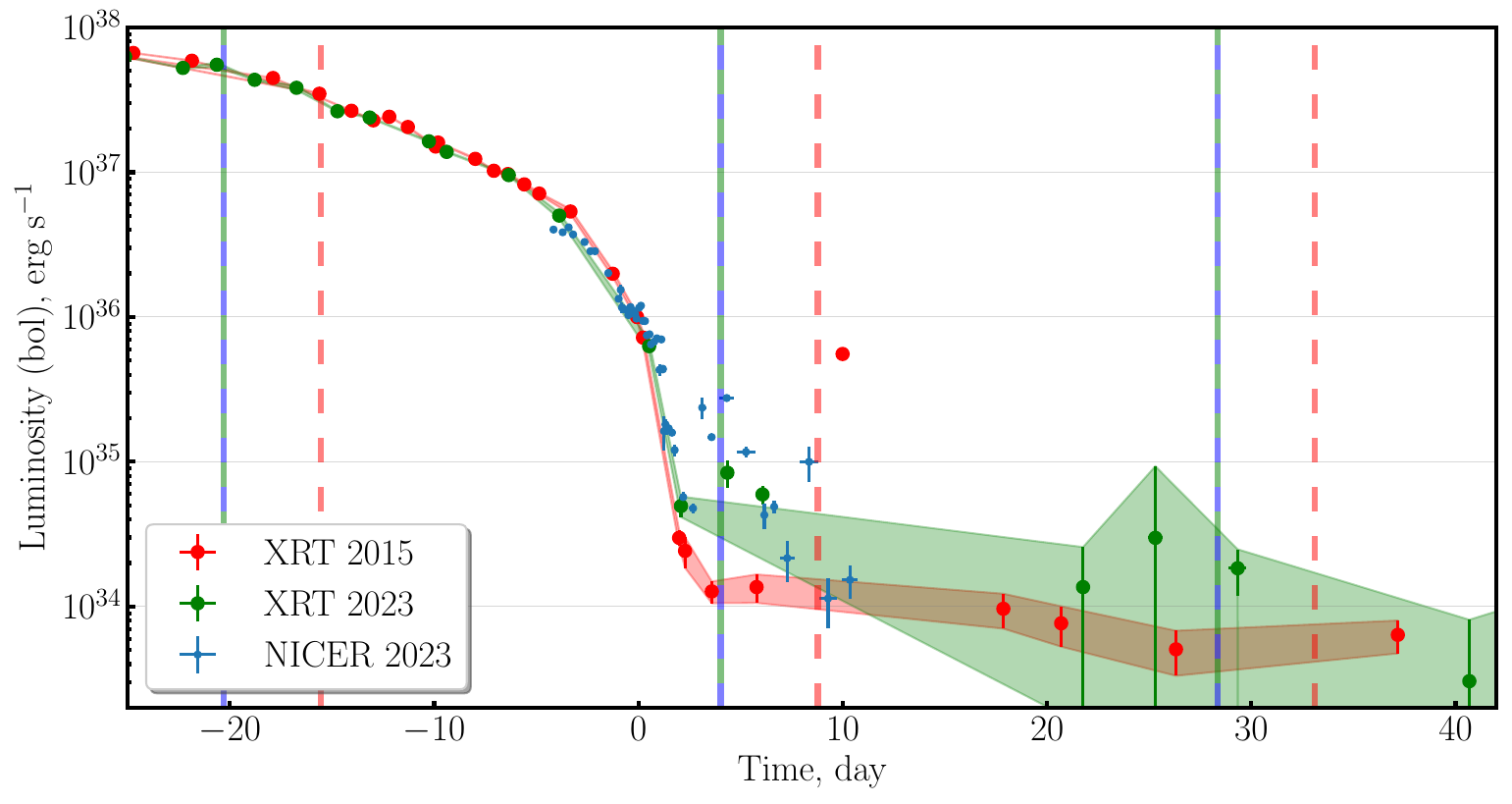}
\caption{Bolometric light curves of \src\ obtained with \textit{Swift}/XRT during the 2015 outburst (red points), and \textit{Swift}/XRT and NICER during the 2023 outburst (green and blue points, respectively). To match the overall shapes of the outbursts, they were shifted in time by MJD 57343.9 and 60072.2 for the 2015 and 2023 outbursts, respectively. 
All luminosities are given assuming the distance of 5.1~kpc.   
The vertical dashed red and green-blue lines indicate times of periastron passages in 2015 and 2023, respectively, adopting orbital ephemeris from \citet{2021MNRAS.503.6045D}. 
Shaded regions highlight the source's overall behaviour during the final stages of the outbursts, with a few outlying points close to the periastron being excluded.
}
 \label{fig:lc-0115}
\end{figure}
%%%%%%%%%%%%%%%%%%%%%%%%%

In reality, both mechanisms (the propeller effect and the DIM) may operate simultaneously, and disentangling their interplay, as well as the roles of physical conditions in the accretion disc and its interaction with the NSs magnetic field, requires both high-quality observational data (i.e. dense monitoring) and detailed theoretical modelling. The best objects for such investigations are X-ray pulsars in binary systems with Be optical companions (Be/X-ray binaries, Be/XRPs). A Be optical companion is a rapidly rotating, non-supergiant massive star that exhibits Balmer emission lines produced by a circumstellar decretion disc. These systems are particularly well suited for such purposes because they are transient sources  that span a very broad range of mass-accretion rates, from deep quiescence to super-Eddington outbursts, thereby allowing us to probe the response of the accretion flow and the NS magnetosphere over orders of magnitude in luminosity.

Two types of outbursts are exhibited by Be/XRPs  \citep{1986ApJ...308..669S}: Type I -- periodic outbursts related to the periastron passage by the NS, with a typical X-ray luminosity of $\sim10^{37}$~\lum, and Type II (or giant) -- very rare events with peak luminosities exceeding several times $10^{37}-10^{38}$~\lum\ \citep[see][for a review]{2011Ap&SS.332....1R}. 
Planning dense monitoring of the transition from the accretion regime to the quiescence requires an accurate prediction of the transition date. 
An exceptional similarity of the shapes of the light curves of the transient Be/XRP \src seen during different outbursts \citep[e.g. the 2015 and 2017 outbursts;][]{2016A&A...593A..16T, 2020A&A...638A.152R} allowed us to organise such an observational campaign with the NICER telescope during a giant outburst from this source in 2023. 

The source \src\ is a well-studied transient XRP with a spin period of $\sim$3.6~s. 
It is most famous for its complex X-ray spectrum modified by five cyclotron resonant scattering features (CRSFs), with a fundamental line around 12\,keV \citep{2000AIPC..510..173H}, providing a direct measurement of the magnetic field of about $10^{12}$~G. 
The orbital parameters of the system, including an orbital period of 24.3 days, are also known  \citep{2010MNRAS.406.2663R,2021MNRAS.503.6045D}. 
The distance to the source is estimated on the basis of the \textit{Gaia} data at $d=5.1$~kpc \citep{2023A&A...677A.134N}. 
Crucially for our study, \src belongs to a small group of XRPs where a rapid drop of the luminosity at the final stages of the outburst was interpreted as a manifestation of the propeller effect \citep{2016A&A...593A..16T}. 
The quiescent state right after the transition was characterised by the thermal spectrum with a typical temperature of 0.5--1.0~keV \citep{2016MNRAS.463L..46W,2017MNRAS.470..126T,2020A&A...638A.152R}. 
The transition time between the two states was estimated to be shorter than a couple of days (based on the gap between two consecutive observations), but has never been time-resolved in any of the XRPs. 
Resolving this transition phase was the primary objective of the observational part of the current work.
 
In this work, we present the results of monitoring observations of the transient XRP \src performed during its transition to the quiescent state at the end of the outburst discovered in late March 2023 by the MAXI all-sky monitor \citep{2023ATel15967....1N,2023ATel15978....1C}.
The results, primarily derived from data collected with the NICER observatory (introduced in Sect.~\ref{sec:data}), are presented in Sect.~\ref{sec:res} and discussed in the framework of the DIM in Sect.~\ref{sec:discussion}. 
We applied the same model to the light curves of several other XRPs, demonstrating fast flux variability at the end of their outbursts.

\section{Data} 
\label{sec:data}

The source \src\ was monitored during the declining phase of its giant outburst from April to July 2023 using the \textit{Swift} observatory. 
The similarity of the outburst light curve to the one observed in 2015 \citep{2016A&A...593A..16T} allowed us to accurately predict the moment of transition of the source to the quiescent state to around MJD~60072.5, which allowed us to trigger our anticipated NICER TOO monitoring campaign several days in advance. 
The main feature of this campaign compared to the previous investigations \citep{2016A&A...593A..16T,2020A&A...638A.152R} was the much higher cadence of observations required to resolve the fast transition to the quiescent state in time. 
In addition, we also re-analysed data from the 2015 outbursts from \src and \srcV \citep{2016A&A...593A..16T}, 2015 and 2022 outbursts from \mbox{SMC~X-2} \citep{2017ApJ...834..209L,2024MNRAS.528.7115C}, and 2017--2018 and 2023 outbursts from \mbox{Swift~J0243.6+6124} \citep{2017ATel10809....1K,2023ATel16076....1P} to further test the model developed to interpret the 2023 outburst light curve of \src and discussed in Sect.~\ref{sec:dim}. 
Below we describe the data and analysis procedures in more detail.

\subsection{NICER}

NICER X-ray Timing Instrument (XTI) is a non-imaging, soft X-ray telescope aboard the International Space Station \citep[ISS;][]{Gendreau2016}. 
It comprises an array of 56 X-ray concentrator optics (XRC) and silicon drift detector (SDD) pairs, operating in the 0.2--12 keV energy band. 
The XTI is characterised with a large effective area of approximately 1900~cm$^2$ at 1.5 keV, an energy resolution of 85 eV at 1 keV, and a time resolution of 300\,ns. 
The scheduling flexibility of NICER makes it an ideal instrument to conduct a high-cadence observational campaign such as the one presented here. 
In particular, we were able to reach the average interval of 2.6 hours between the observations during the particular interval of interest, spanning MJD 60071 to 60074, where the transition of \src to the quiescence was anticipated. 

NICER data were reduced using the {\sc nicerdas} software and {CALDB} version xti20221001 and {\sc heasoft} version 6.32.1, according to the official data analysis threads.\footnote{\url{https://heasarc.gsfc.nasa.gov/docs/nicer/analysis_threads/}} Spectra were extracted with the {\sc nicerl3-spect} pipeline using standard screening criteria. 
To take the contribution of the background into account, the SCORPEON model was applied.  To construct the light curve, 16 observations with ObsIDs 6611030101--11, 6203730132--36, covering MJD 60068--60083, were used. 
The source flux was estimated based on an approximation of the observed spectra using an absorbed power-law model with the hydrogen column density $N_{\rm H}$ fixed at $1.0\times10^{22}$~cm$^{-2}$, following \citet{2016A&A...593A..16T}. 
The NICER spectra were grouped using the optimal binning scheme of \citet{2016A&A...587A.151K} with a minimum of ten counts per grouped bin, and fitted with \textsc{xspec} 12.13.1 \citep{Arnaud1996} using W-statistic \citep{1979ApJ...230..274W}. 
The corresponding fluxes are reported in the 0.5--10~keV range, errors are given at the 1$\sigma$ confidence level if not specified otherwise.

\subsection{Swift}

On longer timescales (and therefore with lower cadence), \src\ was monitored with the X-ray telescope \citep[XRT;][]{2005SSRv..120..165B} onboard the {\it Neil Gehrels Swift Observatory} \citep[\textit{Swift};][]{2004ApJ...611.1005G}. 
Depending on the current source flux, observations were performed either in the photon counting (PC) or windowed timing (WT) mode, followed by the reduction using the online tools \citep[][]{2009MNRAS.397.1177E} provided by the UK Swift Science Data Centre.\footnote{\url{http://www.swift.ac.uk/user_objects/}} 
The source flux in the 0.5--10~keV band was estimated for each individual observation by fitting the corresponding spectrum with the same model that was applied to the NICER data. 
All the spectra were binned to have at least 1 count per energy bin and the W-statistic\footnote{\url{https://heasarc.gsfc.nasa.gov/xanadu/xspec/manual/XSappendixStatistics.html}} was applied \citep{1979ApJ...230..274W}. 
For consistency, we reprocessed the \textit{Swift}/XRT data for all other sources in our sample, following the exact same procedure.

\section{Results} 
\label{sec:res}

The main motivation of the presented monitoring campaign is to construct a high-cadence, densely sampled light curve of the transition of \src\ from the accretion regime to the quiescent state. 
The first step required for interpretation of the obtained high-cadence light curve in the context of accretion physics models is to estimate the accretion rate from the observed X-ray flux. 
There are several potential sources of uncertainty besides the statistical quality of the data: (i) beaming of the emission, (ii) distance to the system, (iii) uncertainty in bolometric correction of the flux (NICER and \textit{Swift}/XRT  cover only the soft X-ray band). 

It was demonstrated by \cite{Markozov2024} that our ignorance of the intrinsic beam pattern of the NS may lead only to a modest deviation of the intrinsic luminosity from the isotropic one, at a level of less than $\sim$20\%. 
Thanks to the \textit{Gaia} data, the uncertainty in the luminosity due to imperfect knowledge of the distance to the Galactic XRPs is now below $\sim$20\% as well. 
Another factor that could potentially affect the shape and normalisation of the light curve is the conversion of the flux in the 0.5--10~keV band to the bolometric flux. 
Given that the spectral shape of XRPs is known to depend on luminosity, we therefore estimated the luminosity-dependent bolometric factor using the broad-band \textit{NuSTAR} observations as described in Appendix~\ref{sec:kbol}. 
The first two sources of uncertainty are not expected to affect the shape of the light curve (only the overall normalisation), whereas the latter, bolometric correction, was carefully accounted for in our analysis.

The resulting bolometric light curve of \src, derived from the \textit{Swift}/XRT and NICER data obtained during 2023 outbursts, is presented in Fig.~\ref{fig:lc-0115} in green and blue, respectively. 
The plot also contains the light curve obtained during the 2015 outburst (shown in red) for comparison. 
Both light curves were shifted in time to match the overall shapes. 
The main result obtained from the NICER monitoring is the detailed coverage of the gap between the subsequent \textit{Swift} observations around the transition (zero time in these coordinates). 
It was found that even after the source luminosity drops below $\sim10^{36}$~\lum, it continues to decrease gradually with an exponential timescale of around 16.5~h. 
The absence of a very sharp drop in the observed flux, which could potentially be associated with the onset of the propeller effect in \src, is one of the main results of our observational campaign.

The observations following both the 2015 and 2023 outbursts show a `meta-stable' state, with a luminosity several times higher than that in quiescence and possibly associated with either low-level accretion or cooling of the accretion-heated NS crust \citep{2016MNRAS.463L..46W}. The main difference between the two outbursts is that the quiescence luminosity after the 2023 outburst was slightly higher than that observed after the 2015 outburst. This difference is most probably related to the periastron passage (shown with vertical dashed red and green-blue lines for the 2015 and 2023 outbursts, respectively), which occurred shortly after the end of the 2023 outburst. Similarly to the \textit{Swift} data, the NICER observations also showed an elevated and variable flux around the time of the periastron passage (see Fig.~\ref{fig:lc-0115}).

\section{Discussion} 
\label{sec:discussion}

The main outcome of the observational component of this research is the first detailed light curve of the transition of \src\ from the accretion to the quiescent state, which occurred over a very short but finite time. 
In this section, we discuss possible physical mechanisms, namely, the propeller effect and the viscous evolution of the disc, that can potentially explain the observed properties of this and other transient XRPs during the final stages of their outbursts.

The propeller effect is the most commonly invoked mechanism to explain the rapid flux decline observed in \src\ and other transient XRPs during the transition to quiescence. 
In its simplest form, accretion onto the NS halts when the mass accretion rate drops below a critical threshold, specifically, the point at which the magnetospheric radius equals the corotation radius \citep{1975A&A....39..185I}. 
Although more sophisticated versions of this model have been proposed \citep[e.g.][]{2017MNRAS.466..175E, 2023MNRAS.520.4315L}, they do not fundamentally alter this general picture.
However, when accretion proceeds through the disc and varies over time, the character of the transition must reflect the evolution of the accretion rate.
The actual temporal evolution during the transition to the propeller state may exhibit a complex dependence governed by the disc–magnetosphere interaction \citep{2005ApJ...635L.165R,2006ApJ...646..304U,2010MNRAS.406.1208D,2017ApJ...851L..34P,2018NewA...62...94R}.
 
In Sect.~\ref{sec:prop}, we explore the propeller mechanism and its implications, focusing on aspects other than the timescale of the transition to the quiescent state. 
The viscous evolution of the disc, and its ability to reproduce the observed decay, is discussed in Sect.~\ref{sec:dim}.

\subsection{Propeller effect}
\label{sec:prop}

Following the definition provided above, the limiting bolometric luminosity marking the onset of the propeller regime can be expressed as follows \citep[see e.g.][]{2002ApJ...580..389C}:
\begin{eqnarray} \label{eq:Llim}
L_{\rm lim} &\simeq &\frac{GM_{\rm NS}\dot{M}_{\rm lim}}{\Rns}  \\ \nonumber
&\simeq & 4 \times 10^{37} k^{7/2} B_{12}^2
P^{-7/3} M_{1.4}^{-2/3} R_6^5 \,\textrm{erg s$^{-1}$},
\end{eqnarray}
assuming dipole magnetic field. 
Here $B_{12}$ represents the NS magnetic field strength at the magnetic poles in units of $10^{12}$~G, $P$ is the pulsar rotational period in seconds, $M_{1.4}$ is the NS mass in units of 1.4$M_\odot$, and $R_6$ is the NS radius in units of $10^6$~cm. 
Here $\dot{M}_{\rm lim}$ corresponds to the condition $R_{\rm cor} = R_{\rm m}=k R_{\rm A}$, where   
\be
R_{\rm cor} = (GM_{\rm NS}P^2/4\pi^2)^{1/3} \approx 1.7 \times 10^8 M_{1.4}^{1/3}\,P^{2/3}\,\textrm{cm}
\ee
is the corotation radius and 
\be \label{eq:ra}
R_{\rm A} = \left(\frac{\mu_{\rm m}^4}{2GM_{\rm NS}\dot M^2}\right)^{1/7} \approx 2.1\times 10^8\, B_{12}^{4/7}\,R_6^{12/7}\,M_{1.4}^{-1/7}\,\dot M_{17}^{-2/7}\,\textrm{cm}
\ee
is the Alfv\'en radius, where $\dot M_{17}$ denotes the accretion rate in units of $10^{17}$~g~s$^{-1}$. 
Here, the relation between the dipole magnetic moment $\mu_{\rm m}$ and the magnetic field strength at the magnetic poles $B$, $\mu_{\rm m} = B \Rns^3/2$ was used. 
The factor $k$ accounts for the discrepancy between the magnetospheric radius and the Alfv\'en radius calculated for spherical accretion ($R_{\rm m}=k\,R_{\rm A}$). 
In the case of disc accretion, $k$ is typically assumed to be 0.5 \citep{1978ApJ...223L..83G}. 

The numerical value of the transitional luminosity to the propeller regime is subject to significant uncertainties related to the strong dependence of Eq.~\eqref{eq:Llim} on poorly constrained parameters, such as the NS radius, the factor $k$, and the magnetic field strength. 
For instance, assuming $k=0.5$, $R_6=1.2$, NS mass of 1.4$M_\odot$, and using other parameters of 4U\,0115+63, namely the spin period $P=3.6$\,s and $B_{12}=1.3$, we obtain $L_{\rm lim} \approx 7.5 \times 10^{35}$\,\lum.
In contrast, assuming the widely used values $R_6=1$ and $B_{12}=1$ results in $L_{\rm lim} \approx 1.8 \times 10^{35}$\,\lum. 
Allowing $k$ being smaller (e.g. 0.3) one can obtain even lower estimates for the transitional luminosity ($L_{\rm lim} \approx 3 \times 10^{34}$\,\lum). 
Moreover, even these estimates may be further uncertain due to possible deviations of the magnetic field structure from a pure dipole configuration and the non-zero magnetic obliquity of the NS, which alters the shape of the equilibrium surface, effectively increasing its radius compared to the classical corotation radius. 
Therefore, conservatively speaking, one might expect the onset of the propeller regime to occur within a broad luminosity range from several $10^{34}$ to $\sim10^{36}$\,\lum.

Some observational properties of transient XRPs have always been considered in the literature as solid manifestations of the onset of the propeller regime, namely the disappearance of pulsations and substantial softening of the emission spectrum. 
In the following, we examine these two observables and discuss their reliability as discriminators of accretion states.

Several XRPs, primarily those with relatively long spin periods, have been observed to exhibit pulsations at luminosities on the order of $10^{33-34}$\,\lum \citep[see Table 2 in][]{2014MNRAS.445.1314R}. 
For these sources, this suggests that matter continues to reach the magnetic poles of the NS even at quiescence luminosities.
Even more interestingly, \citet{2017MNRAS.472.1802R} and \citet{2020A&A...638A.152R} reported pulsations in \src\ several months after its transition to the low state following its 2015 and 2018 outbursts. 
These pulsations could arise either from ongoing low-level accretion or from an inhomogeneous temperature distribution at the NS surface, caused by enhanced heat transport along magnetic field lines from deeper layers.
Therefore, it is clear  that pulsations are observed across all states of transient XRPs, and their presence or absence cannot be reliably used as an indicator of a specific accretion state, including the cessation of accretion.

The substantial softening of emission from the transient XRPs observed in the quiescent state was indicated as an important evidence of the cessation of accretion onto the NS and subsequent thermal emission of the cooling NS surface \citep[see e.g.][]{2016A&A...593A..16T,2016MNRAS.463L..46W,2017MNRAS.472.1802R}. 
However, possibility of a low-level accretion could not be excluded due to unknown physics of spectral formation in such regime. 
For instance, it was demonstrated theoretically that at a very low mass accretion rate the spectral shape is not expected to drastically differ from the blackbody  \citep{1995ApJ...439..849Z}. 
More recently, observations have shown that at low mass accretion rates, XRP spectra undergo a drastic change compared to their classical form, with the emergence of two distinct components: one associated with thermal emission from the deep, nearly isothermal layers of the NS atmosphere, and the other with thermal Comptonisation of radiation originating in the overheated upper layers of the atmosphere \citep{2019MNRAS.483L.144T,2019MNRAS.487L..30T,2021MNRAS.503.5193M,2021A&A...651A..12S}. 
From Fig.~1 in \citet{2019MNRAS.483L.144T} we can see that indeed the spectrum becomes much softer at low mass accretion rates, especially if only soft X-ray band is considered ($<10$~keV), which is the case for the currently available data.
Therefore, similarly to pulsations, the softening of the spectrum alone cannot be considered as a definitive proof of the cessation of accretion.

In this context, a strong argument for ongoing accretion in the quiescent state would be the detection of hard X-ray photons (above $\sim10$~keV) in the broadband spectrum of the source. 
Unfortunately, such deep observations in the low state of \src\ have not been performed yet. 
However, another X-ray pulsar, Swift~J0243.6+6124, was observed in the broad energy band by \textit{NuSTAR} after entering the stable low state with luminosity of the order of $10^{34}$\,\lum  \citep{2020MNRAS.491.1857D}. 
The authors demonstrated that, even in the quiescent state, the source spectrum can be fitted with a classical hard cut-off power-law model, showing a significant contribution to the flux above 20 keV.
The transition from the accretion state to the low state in this source looked very similar to the same transition in \src\ \citep[see Fig. 3 in ][]{2020MNRAS.491.1857D} including a very fast drop of the flux, meta-stable state around $10^{34}$\,\lum with further gradual decrease. 
Clearly, in Swift~J0243.6+6124 this drop was not caused by the propeller effect.

At the same time, very convincing evidence of the cessation of accretion during the quiescent state in \src\ was recently reported by \cite{Xiao2025}. Using deep \textit{XMM-Newton} observations obtained after the 2023 outburst, the authors demonstrated a complete absence of the red-noise component in the power density spectrum, which is generally observed in accreting systems and originates from the stochastic nature of viscosity in the accretion disc \citep{1997MNRAS.292..679L}. However, since this result was obtained in deep quiescence, a few months after the outburst ended, it does not explain the properties observed immediately around the transition, namely, the flux decay timescale from the accretion state and the origin of the elevated luminosity during the meta-stable state. In the following, we propose an alternative model that explains all the observational properties of \src\ discussed above.

%%%%%%%%%%%%%%%%%%%%%%%%%%%%%%%%%%%%%%%%%%%%%%%%%
\subsection{Thermal-viscous disc instability}
\label{sec:dim}
%%%%%%%%%%%%%%%%%%%%%%%%%%%%%%%%%%%%%%%%%%%%%%%%%

In this section, we discuss an alternative to the propeller mechanism that can also produce a sharp change in the observed luminosity between the accreting and quiescent states, namely, the viscous evolution of the accretion disc.
We suggest that type~II outbursts in Be/XRPs are driven by the DIM mechanism, which has been extensively studied in other classes of accreting systems.
Within the DIM framework, during the decay phase of an outburst, the accretion disc consists of two physically distinct regions. 
The inner region is hot and almost completely ionised, has a high viscosity ($\alpha \gtrsim 0.1$), and can be approximately described by the standard Shakura-Sunyaev disc model \citep{1973A&A....24..337S}. 
The outer region is cooler, with hydrogen being predominantly in the atomic form, which results in a presumably much lower viscosity ($\alpha \sim 0.01$) and a strongly reduced mass accretion rate through this part of the disc.

Below, we consider a time-dependent model of the viscously evolving $\alpha$-disc and demonstrate how the steepening of the light curve occurs in the DIM. 
During the brightness decay phase, the instantaneous mass accretion rate is approximately proportional to the total mass contained in the hot inner region of the disc.
Consequently, the rate of luminosity decay is governed by how rapidly the size of this hot region decreases, that is, by the inward propagation of the cooling (transition) front separating the hot and cold zones.

Around magnetised NSs, the inner radius of the ionised zone of the accretion disc is truncated at the magnetospheric radius, while the cooling front defines its outer edge. 
This causes the hot part of the disc to shrink gradually from both the inner and outer edges. 

In XRPs, irradiation of the accretion disc by the central NS is a key ingredient of the decay. 
At sufficiently high luminosities, the incident X-ray flux maintains a large ionised (hot) region, and the position of the cooling front is primarily controlled by irradiation rather than by local viscous heating. 
The irradiation flux,  $Q_{\rm irr} = {C_{\rm irr}}\, L /({4\pi R^2}) $, where $L$ is the central luminosity and $C_{\rm irr}$ is the irradiation parameter, exceeds the viscous flux, $Q_{\rm vis} = 3 {GM\dot M }/( {8\pi} R^3)$, only at sufficiently large radii. 
Thus, the decay naturally consists of two stages: an irradiation-controlled stage at higher luminosities and a purely viscous stage at lower luminosities. 
Consequently, the typical evolution of an X-ray transient begins with an irradiation-controlled stage, ending when the cooling front reaches the radius where the irradiation flux drops below the intrinsic viscous flux.
This radius, $R_{\rm irr,min}$, depends on the irradiation parameter of the disc $C_{\rm irr}$ (see Eq.~(41) of \citealt{Suleimanov+2007}):
\be\label{eq:cirr}
  C_{\rm irr} =\eta \Psi(\theta) \left( \frac{dz}{dR} - \frac{z}{R} \right)\approx \frac{1}{12}\,\eta\, \Psi(\theta)\,\frac{z}{R}\, ,
\ee
where $z$ is the disc half-thickness at the given disc radius $R$, \mbox{$\eta\sim$0.1--0.5} is a thermalisation efficiency, and $\Psi(\theta)\approx1$ is an angular flux distribution of the XRP. 
The approximate equality in Eq.~\eqref{eq:cirr} arises from applying the accretion $\alpha$-disc solution near the outer radius of the hot zone \citep{Tavleev+2023}: 
\be \label{eq:zr}
  \frac{z}{R} \approx 0.03\,M_{1.4}^{-13/36}\dot M_{16}^{1/6}\,\left(\frac{\alpha}{0.1}\right)^{-1/9}
  R_9^{1/12}.
\ee
Relation \eqref{eq:zr} differs from the standard Shakura-Sunyaev solution for the zone dominated by free-free opacity (zone~C) because the Rosseland plasma opacity at relatively low temperatures ($\sim$ $10^4$\,K) is better described by the dependence $\kappa_{\rm R} \propto \rho T^{-5/2}$ rather than Kramers' law ($\kappa_{\rm R} \propto \rho T^{-7/2}$).
Thus, we can estimate
\be
C_{\rm irr} \approx 6\times 10^{-4}\, \left(\frac{\eta}{0.2}\right)\,\Psi(\theta)\,M_{1.4}^{-13/36}\dot M_{16}^{1/6}\,\left(\frac{\alpha}{0.1}\right)^{-1/9}
  R_9^{1/12}.
\ee
We note that this estimate is obtained for direct irradiation of the disc by the central source, neglecting the influence of the transition layer between the disc and magnetosphere, scattering in the accretion curtain, and scattering in the extended disc atmosphere. 
The influence of these components may be significant, as scattering in the accretion curtain can increase the irradiation, while a hot transition layer, which is thicker than the disc, can, conversely, reduce the irradiation through shielding.

The radius $R_{\rm irr,min}$ can be found from the equality $Q_{\rm visc} = Q_{\rm irr}$ or
\be
\label{eq:Qirr_Qvis}
\frac{3}{8\pi} \frac{GM}{R_{\rm irr,min}^3} {\dot M} = \frac{GM}{\Rns} {\dot M} \frac{C_{\rm irr}}{4\pi R_{\rm irr,min}^2}.
\ee
Finally,
\be
R_{\rm irr,min} \approx 1.5\, \Rns \,C_{\rm irr}^{-1} \approx 1.5\times 10^{9}\, R_6 \,C_{\rm irr,-3}^{-1}\,\textrm{cm},
\ee
where $C_{\rm irr, -3} = C_{\rm irr}/10^{-3}$.
This radius is larger than the corotation radius and irradiation does not affect the dynamics of the viscous disc in 4U\,0115+63 near the corotation radius unless the irradiation parameter is as large as $C_{\rm irr} \approx 3 \times 10^{-3}$.

\begin{figure}
\includegraphics[width=0.99\linewidth]{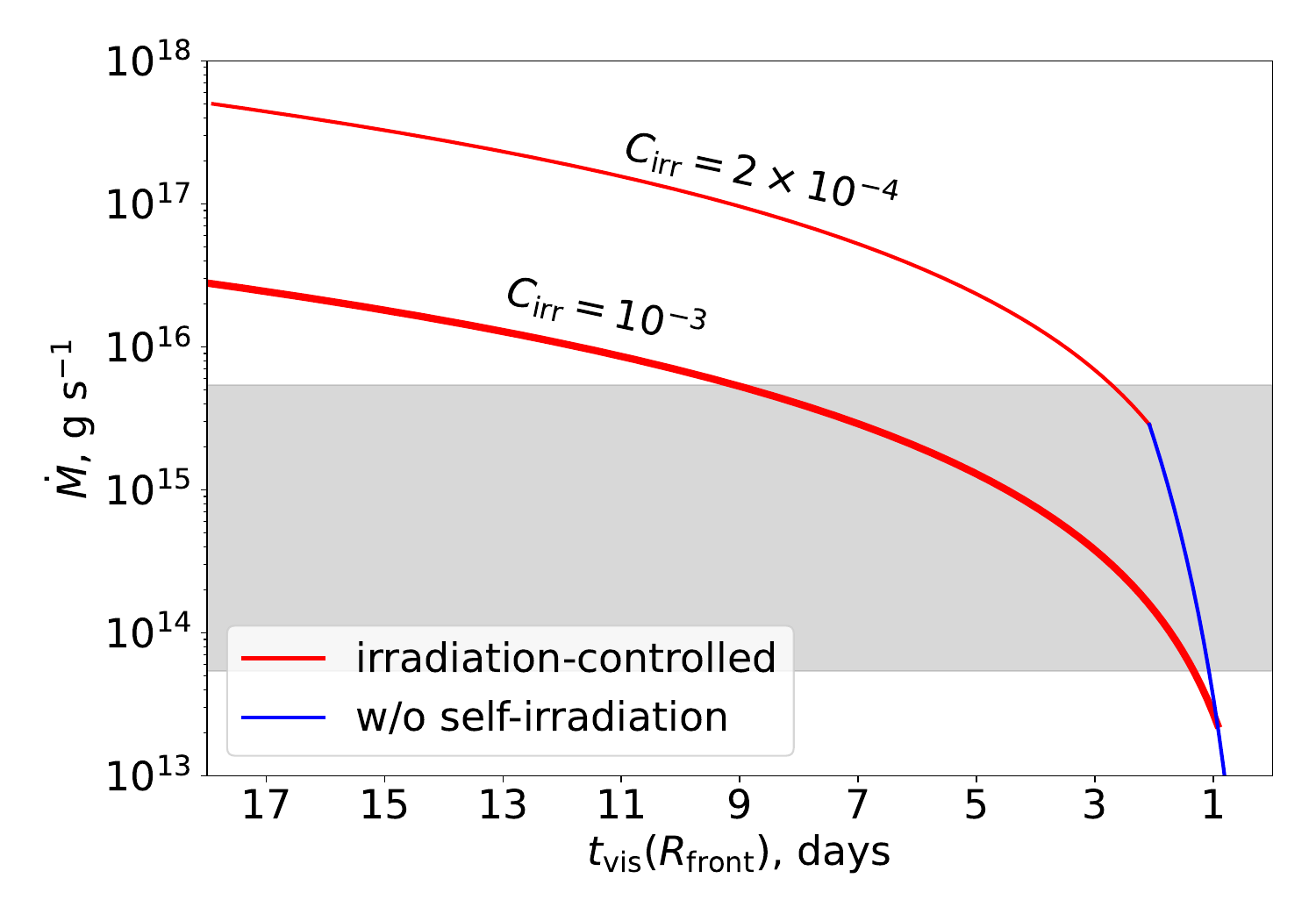}
\caption{Mass accretion rate versus the viscous time at the hot-zone radius for disc evolution with irradiation (red) and without irradiation (blue). Model parameters are $\alpha = 0.3$ and $C_{\rm irr} = 2\times10^{-4}$ (thin line) and $C_{\rm irr} = 10^{-3}$ (thick line). 
The grey horizontal band indicates the conservative range of mass accretion rates corresponding to the possible transition of \src to the propeller regime, i.e. the luminosity interval $10^{34}$–$10^{36}$\,\lum (see Sect.~\ref{sec:prop}).}
\label{fig:decay_times}
\end{figure}

The characteristic timescale of the accretion-rate evolution in the disc is given by the viscous time evaluated at the outer edge of the hot zone (i.e. at the cooling-front radius). 
As the radius of the hot zone decreases, the flux decay timescale shortens, leading to the rapid drop observed in the light curve. 
This behaviour is illustrated in Fig.~\ref{fig:decay_times}, where we show the accretion rate as a function of the viscous time $t_{\rm vis}(R_{\rm front}) = R_{\rm front}/v_{\rm vis}$. The radial, or viscous, velocity $v_{\rm vis} = \alpha \, (z/R)^2 \sqrt{GM/R_{\rm front}}$ is evaluated at the hot-zone outer radius.
This radius is determined from $Q_{\rm irr} = \sigma_{\rm SB}\, T_{\rm crit,irr}^4 $ for $R_{\rm front}>R_{\rm irr,min}$, and 
$Q_{\rm vis} = \sigma_{\rm SB}\, T_{\rm front}^4 $ otherwise, where $\sigma_{\rm SB}$ is the Stefan-Boltzmann constant. For simplicity, 
we adopted a single representative temperature at the hot zone boundary,
$T_{\rm front} = T_{\rm crit,irr} \approx 7000$~K~\citep[see the discussion in][]{Tavleev+2023}. 
The qualitative picture of Fig.~\ref{fig:decay_times} is confirmed by the numerical solution of the disc viscous-evolution equation, as demonstrated for the case of the outburst evolution of a low-mass X-ray binary \hbox{Aql\,X-1}  by \citet{Lipunova+2022} and \citet{2024ApJ...961..252C}.  
In the same figure, the grey horizontal strip indicates a plausible (conservative) range of mass accretion rates for the transition of 4U~0115+63 to the propeller state (see Sect.~\ref{sec:prop}).

In the following, we present the first numerical fits for the light curves at the late stages of outbursts in several transient XRPs. 
In addition to \src\ in 2015 and 2023, we also consider several XRP outbursts observed by \textit{Swift}/XRT, which are listed in Table~\ref{tab:ns_par}. 
The spectral modelling is described in Appendix~\ref{sec:kbol}. 
The bolometric fluxes were converted to $\dot M$, assuming accretion efficiency 20\%. 
Theoretical $\dot M(t)$ is calculated using the viscous-disc-evolution code \freddi \citep{Lipunova-Malanchev2017},\footnote{\url{https://github.com/hombit/freddi/blob/master/Readme.md}} with some modifications described in Appendix~\ref{sec:freddi_details}. 
It is assumed that there is no magnetic gating effect at $R_{\rm m}$, which means that all matter that reaches the inner boundary of the disc falls onto the NS. 
The viscous torque at the inner disc  radius is set to $0.5^{7/2}\, \mu_m^2/R_{\rm cor}^3$ \citep[see][sect.~2.3 of]{Lipunova+2022}.

The main input parameters controlling the disc evolution rate are the turbulent viscosity parameter $\alpha$ and the irradiation parameter $\tilde C_\mathrm{irr}$.  
Parameter $\tilde C_\mathrm{irr}$ serves as a proxy for the irradiation parameter $C_{\rm irr}$, as explained in Appendix~\ref{sec:freddi_details}, and has a comparable magnitude.
It was previously established that, for a model of an evolving $\alpha$-disc such as \freddi, there is a degeneracy and a positive correlation between the parameters $\alpha$ and $C_{\rm irr}$.
In the case of HMXBs (including XRPs), the irradiation parameter $C_{\rm irr}$ cannot be reliably constrained from the observed optical flux; therefore, all our fits and conclusions must account for this degeneracy.

\begin{figure*}
 \centering
\includegraphics[width=0.43\linewidth]{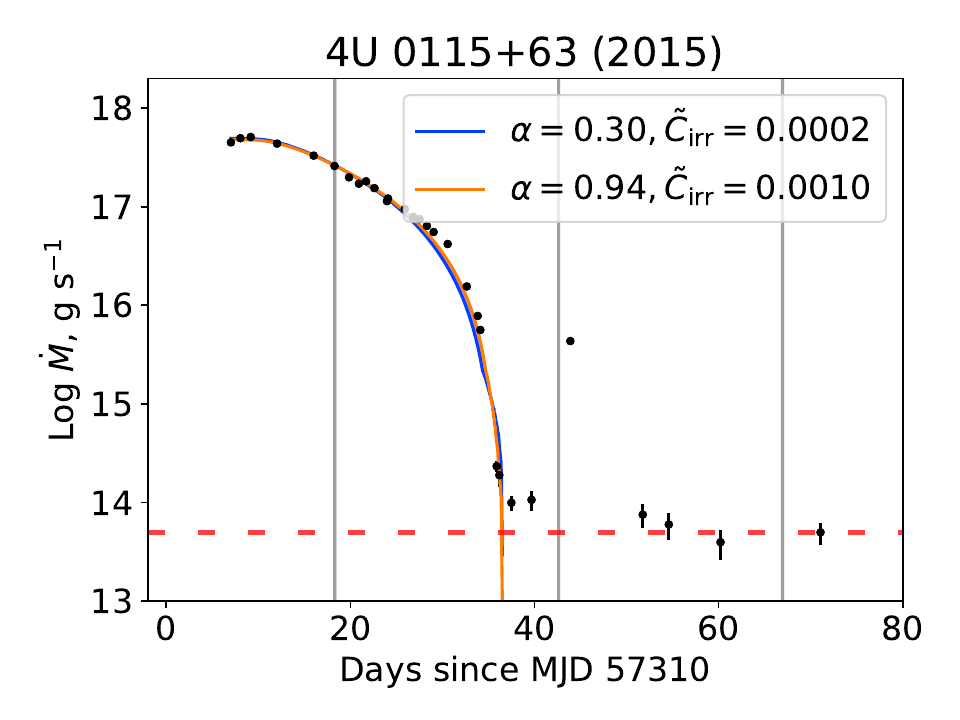}
 %\hfill
 \includegraphics[width=0.43\linewidth]{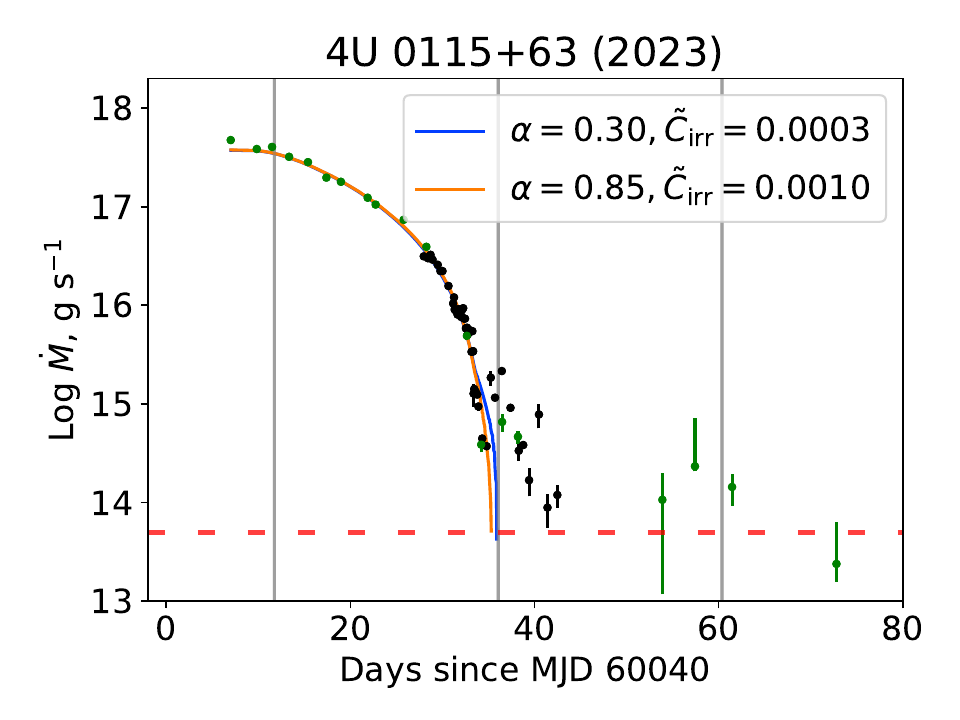}
\includegraphics[width=0.43\linewidth]{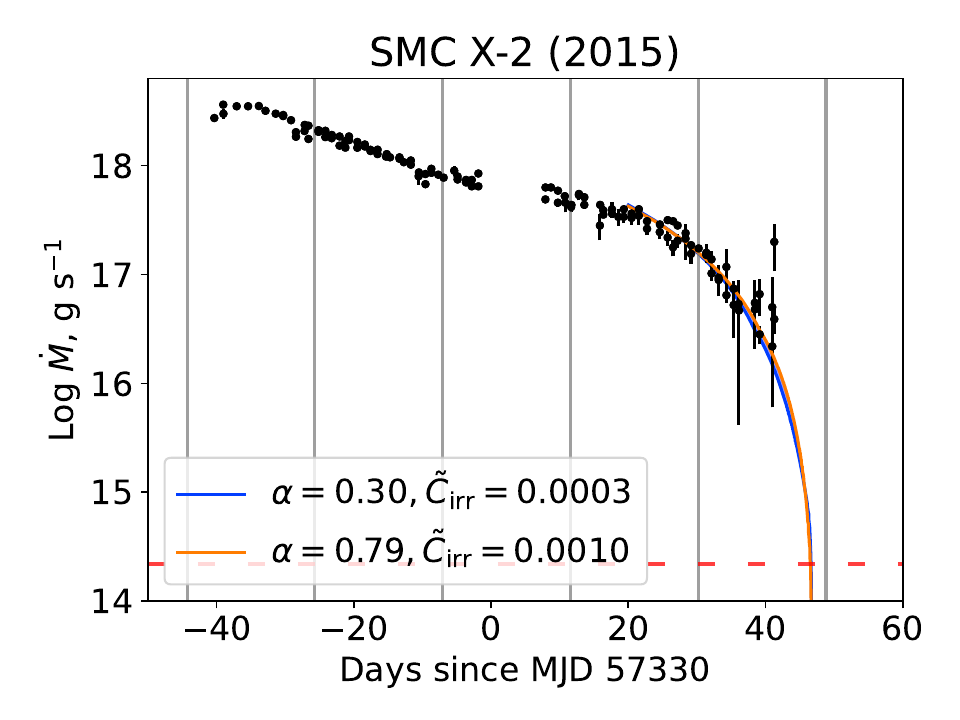}
% \hfill
\includegraphics[width=0.43\linewidth]{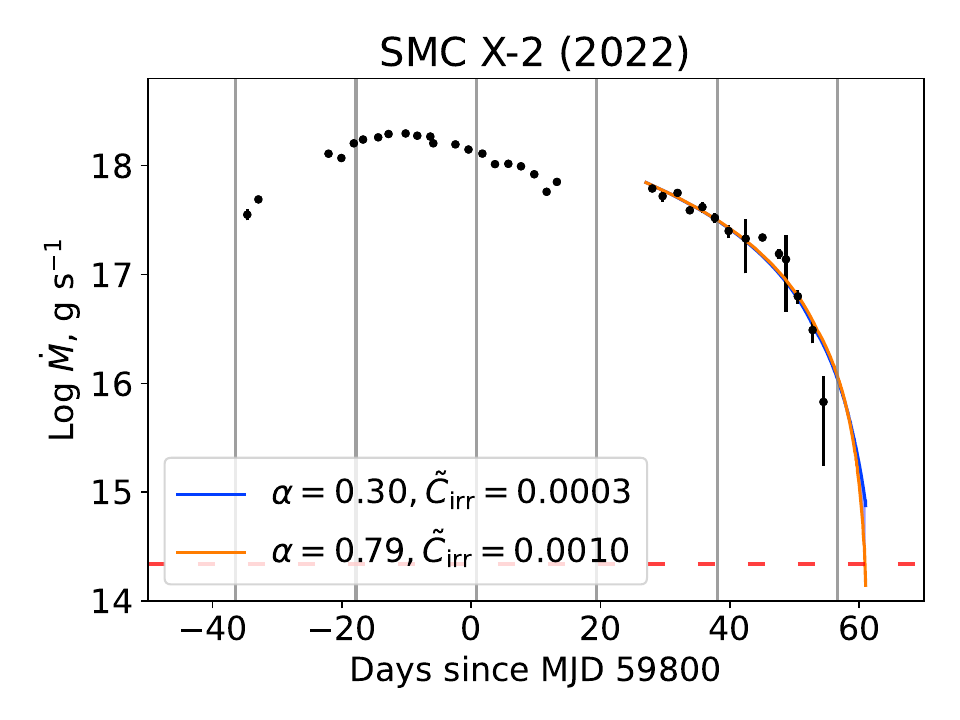}
 \includegraphics[width=0.43\linewidth]{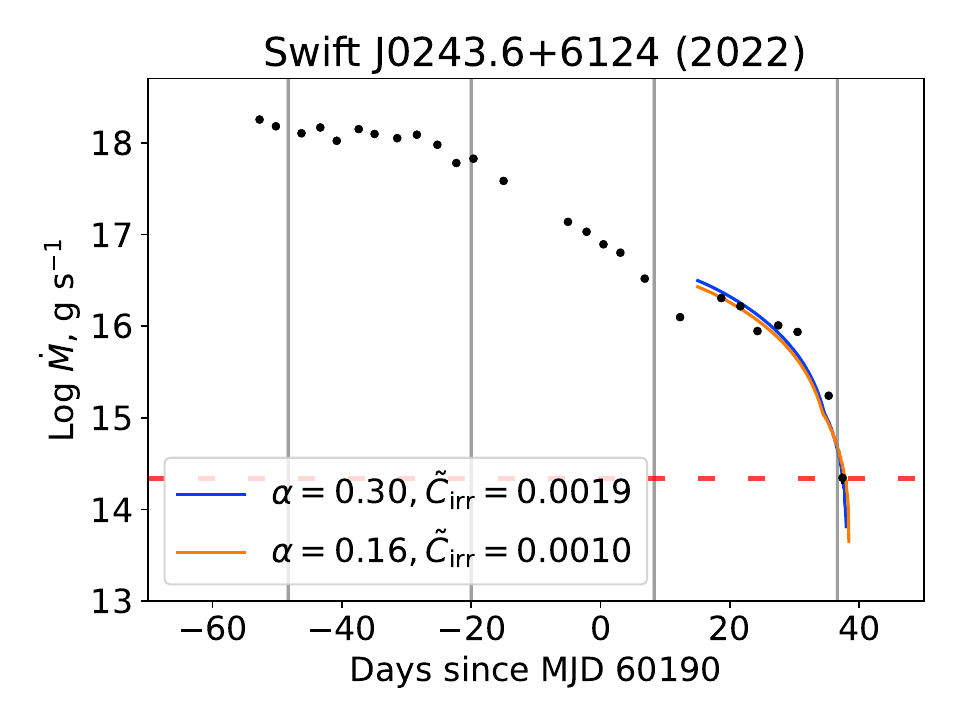}
% \hfill
\includegraphics[width=0.43\linewidth]{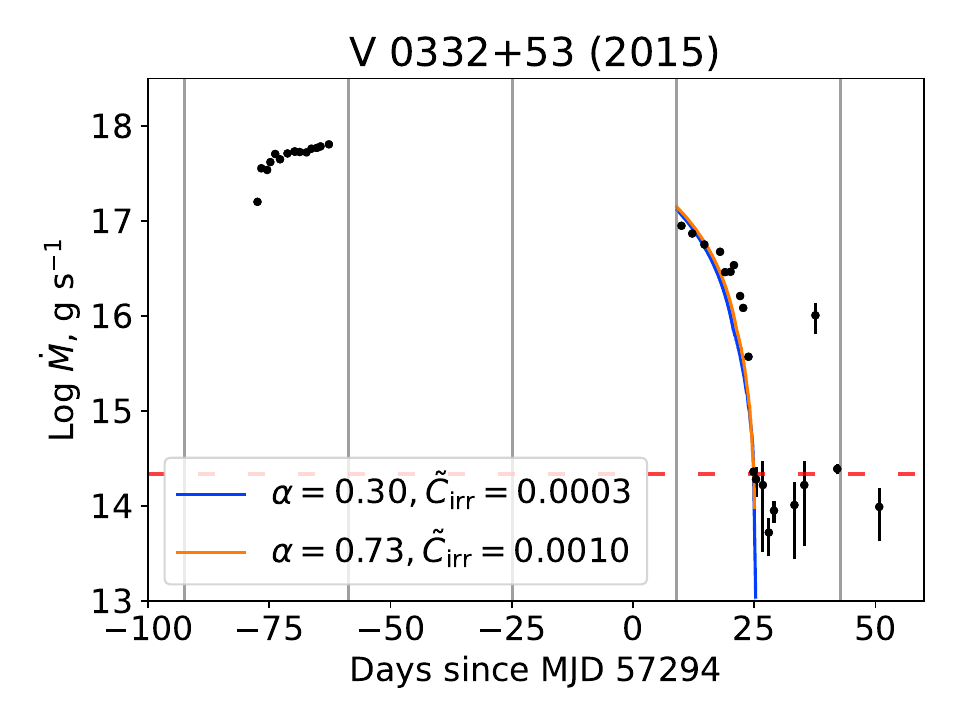}
\caption{Fits with the \freddi code to the light curves during the final stages of XRP outbursts, assuming either a fixed viscosity parameter $\alpha = 0.3$ (blue) or a fixed irradiation parameter $\tilde{C}_{\rm irr} = 0.001$ (red). 
The best-fitting values of the free parameter ($\tilde{C}_{\rm irr}$ or $\alpha$, respectively) are indicated in the legends. Vertical grey lines mark the periastron passages. The horizontal dashed red lines represent the expected accretion rate in the cold disc state, calculated using Eq.~\eqref{eq:L_cold}, with $A=0.057$, $k=0.5$, and $B$ from Table~\ref{tab:ns_par}.}
\label{fig:models_freddi}
\end{figure*}

We therefore present two alternative fits to the observed light curves.
In the first case, we fixed the parameter  $\tilde C_\mathrm{irr} = 10^{-3}$ and determined the best-fit values of the turbulent parameter $\alpha$.
In the second case, we fixed the viscosity parameter at $\alpha = 0.3$ and determined the best-fit values of the irradiation parameter $\tilde C_\mathrm{irr}$. 
In both cases, we fixed the parameter $k = R_{\rm m}/R_{\rm A} = 0.5$.\footnote{The results depend only weakly on the value of $k$.}
The resulting fits are shown in Fig.~\ref{fig:models_freddi} for the first (red curves) and second (blue curves) cases. 

It is evident that the proposed model reproduces the observed light curves during the final stages of the giant outbursts in all considered sources without invoking the propeller mechanism. At the same time, the currently available data are insufficient to completely exclude possible effects of magnetic gating, which is expected to operate over a broad luminosity range from several $10^{34}$ to $\sim$$10^{36}$\,\lum (see Sect.~\ref{sec:prop}).
We note that the flux decline in the studied systems is relatively slow, which implies either a low value of $\alpha$ or a strong irradiation effect, as both factors are known to slow the luminosity decay during non-stationary disc accretion.

%%%%%%%%%%%%%%%%%%%%%%%%%%%%%%%%%%
%%%%%%%%%%%%%%%%%%%%%%%%%%%%%%%%%%

% %%%%%%%%%%%%%%%%%%%%%%%%%%%%%%%%%
% %%%%%%%%%%%%%%%%%%%%%%%%%%%%%%%%%

%%%%%%%%%%%%%%%%%%%%%%%%%%%%%%%%%
\begin{table}
\centering
\caption{Input NS parameters.}
\begin{tabular}{c c c c}
  \hline  \hline
Source & $P$ (s) & $B$ ($10^{12}$ G)  
  & $d$ (kpc) \\ 
  \hline
4U\,0115+63 & 3.60 & 1.3 & $5.8_{-0.5}^{+0.8}$\\ 
SMC\,X-2  & 2.40 & 3.0& $62.4\pm0.9$\\ 
Swift\,J0243.6+6124 & 9.86 & 3.0$^a$ & $5.2\pm0.3$\\ 
V\,0332+53  & 4.40 & 3.0 & $5.6^{+0.7}_{-0.5}$ \\ 
\hline
\end{tabular}
\tablefoot{
$^a$The estimate for the dipole component of the magnetic field is taken from \cite{2020MNRAS.491.1857D}. For the other sources, the magnetic field is inferred from the cyclotron line energy \citep{2019A&A...622A..61S}.
}
\label{tab:ns_par}
\end{table}
%%%%%%%%%%%%%%%%%%%%%%%%%%%%%%%%%

 %%%%%%%%%%%%%%%%%%%%%%%%%%%%%%%%%%%%%%%%%%%%%%
\subsection{Nature of the low-luminosity quasi-stable state}
%%%%%%%%%%%%%%%%%%%%%%%%%%%%%%%%%%%%%%%%%%%%%%%

As mentioned above (see Sect.~\ref{sec:prop}), several authors have found that the rapid luminosity decline observed in sources such as \src, \srcV, and Swift~J0243.6+6124 abruptly halts at a level of approximately $10^{34}$\,\lum, after which it continues to decrease gradually over timescales of several months \citep{2016MNRAS.463L..46W,2017MNRAS.472.1802R,2020MNRAS.491.1857D}. 
This behaviour is naturally explained within the framework of the DIM discussed in the previous section.

The presence of a slowly decaying plateau phase at a luminosity of $\sim 10^{35}$\,\lum in the light curve of another transient XRP, \gro, was previously reported by \cite{2017A&A...608A..17T}. 
The authors argued that, in XRPs with a large magnetospheric radius, i.e. with a strong magnetic field, the cooling front may reach the inner disc radius at relatively high mass accretion rates, before the propeller luminosity is reached. 
Once this accretion rate is attained, the rapid decline in the source intensity halts, and the system transitions to a regime of stable accretion from a `cold' (recombined) disc.
Similar behaviour was later confirmed in several other XRPs \citep[see e.g.][]{2017MNRAS.470..126T,2018A&A...620L..13R, 2019A&A...621A.134T}.

Below, we examine whether the same mechanism can account for the observed cessation of the rapid luminosity decline in \src, \srcV, and \sw, the absolute luminosity level at which it occurs, and the subsequent plateau. 
In the case of \smc, the presence of such a low-luminosity plateau cannot be excluded due to limited statistics.

\begin{figure}
\centering
\includegraphics[width=1.05\linewidth]{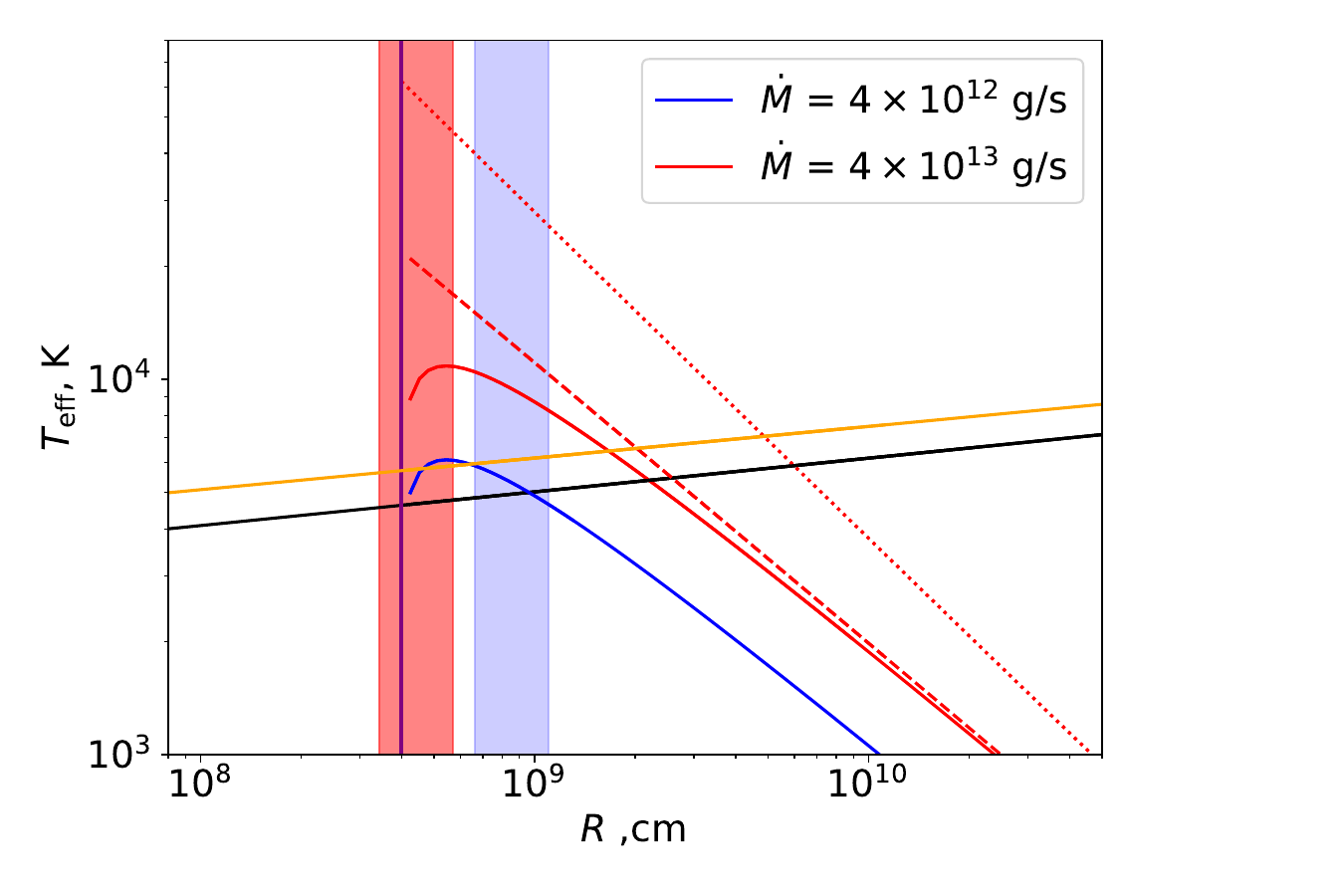}
\caption{Radial distribution of the effective temperature in the $\alpha$-disc. The distributions for two accretion rates and various inner boundary conditions are shown. Two solid curves (red and blue) correspond to the case $\beta = 1$, while dashed red  line is for $\beta = 0$, and the dotted red line is for $\beta <0$. Straight, nearly horizontal orange and black lines indicate the critical temperatures for the hot stable regime (upper) and the cold stable regime (lower). The vertical thick blue  line is the corotation radius for \src. The coloured stripes (for two different mass accretion rates) show the magnetospheric radius as given by Eq.~\eqref{eq:ra} for $k = 0.15$--$0.25$.}
\label{fig:Teff_R}
\end{figure}

To estimate the luminosity at which the source transitions to a quasi-stable regime of accretion from a cold disc, we follow the simplified approach proposed by \citet{2017A&A...608A..17T}. 
Under this, the critical luminosity below which the disc enters a cold state is obtained from the condition of equality of the radius of the disc with $T_{\rm eff} = 7000$\,K to the magnetospheric radius, $R(T_{7000})=k\,R_{A}$: 
\be
\label{eq:L_cold}
  L_{\rm cold} \simeq 8\times 10 ^{33}A^{-7/13}\,k^{21/13}\,M_{1.4}^{3/13}\,R_6^{23/13}\,B_{12}^{12/13}\,
  T_{7000}^{28/13}
  %\left(\frac{T}{7000}\right)^{28/13}
  \,\textrm{erg s$^{-1}$}.
\ee
The coefficient $A$ is a dimensionless parameter that depends on the parameter $\beta$:
\begin{equation}
A =
\begin{cases}
0.057\,\beta^{-6}, & \beta \ge \sqrt{3}/2,\\[6pt]
1 - \beta, &  0.5 < \beta < \sqrt{3}/2.
\end{cases}
\label{eq:A_def}
\end{equation}
The parameter $\beta$ describes the inner boundary condition associated with angular momentum conservation in an accretion disc \citep[see e.g.][]{1983bhwd.book.....S}. 
The case $\beta = 1$ corresponds to zero viscous stress at the inner edge of the disc. 
The case $\beta = 0$ implies a positive viscous stress at the inner edge, with the corresponding torque exactly compensating for the angular momentum removed from the disc by matter accreting onto the star. 
Negative values of $\beta$ are also possible in systems hosting magnetised NSs \citep[e.g.][]{1974MNRAS.168..603L,1993ApJ...402..593S,Kluzniak-Rappaport2007,2012MNRAS.420..416D,Lipunova+2022}.

These cases are illustrated in Fig.~\ref{fig:Teff_R}, which shows the radial profiles of the disc effective temperature for two accretion rates, indicated by different colours in the legend. Red and blue solid lines correspond to $\beta = 1$ and  demonstrate the characteristic drop due to the requirement that the viscous stress and, consequently, the generated heat and  temperature  are zero at the inner radius. The dashed red  line shows the case of $\beta=0$, and dotted line to $\beta =-4.66$. 
The negative $\beta$ was estimated by assuming that the viscous torque at the inner disc is equal to $0.1\, \mu_m^2/R_{\rm cor}^3$~\citep[see eq. (7) in][]{Lipunova+2022}.
The vertical solid line marks the corotation radius, while the semi-transparent shaded regions indicate the magnetospheric radius for the two accretion rates, assuming $k = 0.15$–$0.25$.\footnote{Such low values of $k$ may be expected for large inclination angles between the magnetic and spin axes of the neutron star \citep{2018A&A...617A.126B}.}
The inclined solid orange and black lines show critical minimum and maximum effective temperature in the $\alpha$-disc in the stable  hot and stable cold regimes,  respectively, calculated following \citet{Tavleev+2023}.  In the DIM framework,  the accretion rate between these two values leads to unstable accretion.  For the scenario of accretion from the cold disc, the effective temperature should be less than the lower limit (the black line)  at any radius. 
 
The higher accretion rate, $4\times 10^{13}$~g\,s$^{-1}$, corresponds to the luminosity of \src in the quasi-stable plateau observed right after the outburst ends.  We see that for any $\beta$, the inner portion of the disc, adjacent to the corotation radius, is thermally unstable. A ten-fold weaker accretion would be a marginally acceptable value for the cold disc scenario with $\beta =1$ in \src (the blue curve). 
However, the inner radius of the disc for such accretion rate is beyond the corotation radius, as can be seen from the figure.
 
Apparently, these discrepancies present certain difficulty for a cold-disc scenario  in the case of  \src. A similar conclusion holds for V~0332+53 and \sw, which also have a short spin period. At the same time, in the latter case, the presence  of accretion during this state was evident \citep{2020MNRAS.491.1857D}. The nature of the post-outburst plateau might be approached in a different way.  

The agreement between the observed transition of the source into the quasi-stable low-luminosity state and the predicted value of $L_{\rm cold}$ given by Eq.~\eqref{eq:L_cold} is a conspicuous fact.  This is a strong  indication that there is a residual accretion from a cold disc. To mitigate the difficulties described above, one could propose that structure of the disc is strongly influenced by the magnetic field or become optically thin so that it does not suffer from thermal instability in the unstable region (between the  orange and black lines in Fig.~\ref{fig:Teff_R}).  
The construction of a detailed model is beyond the scope of the present
paper and will be published elsewhere.

%========================================
%========================================
\section{Summary}
\label{sec:sum}

The results of our work are summarised below:
\begin{itemize}
\item For the first time, the very final stages of a giant outburst from a transient
Be/X-ray pulsar were observed with high time cadence, allowing the transition
timescale to the quiescent state to be constrained.

\item The obtained light curve was successfully modelled using viscous disc
evolution, with the hot zone decreasing in size and without the need to invoke
the propeller effect. The corresponding disc viscosity parameter and irradiation parameter have reasonable values. At the same time, we cannot conclusively rule out some influence of magnetospheric interaction during the outburst decay.

\item Low-luminosity plateaus observed in several transient Be/XRPs 
immediately after their giant outbursts can be interpreted as a transition to
accretion from a cold disc. The observed luminosities of these transitions in \src, \srcV and \sw agree well with the estimates obtained from Eq.~\eqref{eq:L_cold}. 
However, this scenario is valid only if the propeller effect does not set in at a limiting luminosity expected to be approximately an order of magnitude higher. The eventual cessation of accretion observed in deep quiescence may indicate either the onset of an efficient propeller regime at sufficiently low accretion rates or complete depletion of the disc. Clarifying the mechanism responsible for this final transition requires further observations of XRPs in the meta-stable state.
\end{itemize}

%========================================
%========================================
\begin{acknowledgements}
We are grateful to the anonymous referee for their careful reading of our manuscript and for the valuable comments and constructive suggestions. The authors thank Sergio Dzib for the comments on the manuscript. We acknowledge support from the Academy of Finland grants 349144, 349373, and 349906 (SST, JP), the German Academic Exchange Service (DAAD) travel grant 57525212 (VFS), the Vilho, Yrjö and Kalle Väisälä Foundation of the Finnish Academy of Science and Letters (SVF, SST), the German Research Foundation (DFG) grants \mbox{LI 4610/1-1} (GL) and \mbox{WE 1312/53-1} (VFS), UKRI Stephen Hawking fellowship (AAM), the Jenny and Antti Wihuri Foundation grant 00240331 (AS).
This research was supported by the  Research Council of Finland Centre of Excellence in Neutron-Star Physics (grant 374064).
\end{acknowledgements}

\bibliographystyle{aa}
\bibliography{allbib}

%%%%%%%%%%%%%%%%% APPENDICES %%%%%%%%%%%%%%%%%%%%%

\begin{appendix}

\section{Bolometric correction}
\label{sec:kbol}

\begin{table*}%[h]
\centering
{\small
\caption{Observed fluxes in the 0.5--79 keV and 0.5--10 keV bands and the corresponding bolometric correction factors $K_{\rm bol}$ for the analysed sources.}
\label{tab:obsids}
\begin{tabular}{lccccc}
\hline
\hline
Object & ObsID & $F_{0.5-79}$ & $F_{0.5-10}$ & $K_{\rm bol}$ \\
               &                & \multicolumn{2}{c}{($10^{-10}$~\flux)} &  \\

\hline

 \sw       
       & 90501310002 & $0.053^{+0.004}_{-0.003}$ & $0.026\pm0.001$  & $2.0^{+0.2}_{-0.1}$\\
       & 90401308002 & $13.2\pm0.1$& $5.45\pm0.02$ & $2.42\pm0.02$\\
       & 90401334002 & $90.8\pm0.2$ & $31.5_{-0.2}^{+0.1}$ & $2.88^{+0.02}_{-0.01}$\\
       & 90302319002 & $99.1_{-0.3}^{+0.2}$ & $36.1\pm0.2$  & $2.74\pm0.02$\\
       & 90901321002 & $336\pm1$& $118\pm1$ & $2.85\pm0.03$\\
       & 90302319008 & $818\pm2$& $294\pm1$ &$2.78\pm0.02$ \\
\hline
\src        
       & 90902316004 &  $51.4\pm0.1$ & $25.94^{+0.04}_{-0.03}$ & $1.982\pm0.005$\\
       & 90102016004 & $127.0\pm0.2$ & $57.2\pm0.1$ 
       & $2.220\pm0.005$\\
       & 90102016002 & $186.5\pm0.3$ & $84.3\pm0.2$
       & $2.212\pm0.006$\\
       & 90902316002 & $221.7\pm0.2$&$100.0\pm0.2$ & $2.217\pm0.005$\\
\hline
       \smc & 90101017002 & $4.40\pm0.02$ & $1.55\pm0.01$ &
       $2.84\pm0.02$\\
       & 90801319002 & $6.17\pm0.02$ & $2.39\pm0.01$ &
       $2.58\pm0.01$\\
       & 90102014004 & $6.41\pm0.02$ & $2.51\pm0.01$ &
       $2.55\pm0.01$\\
       & 90102014002 &         $13.41\pm0.03$ &        $6.13\pm0.02$ & $2.188\pm0.009$\\

\hline
\srcV  & 90202031004 & $9.02^{+0.05}_{-0.04}$&         $2.76\pm0.05$ &
       $3.27\pm0.06$\\
       & 90202031002 & $10.9\pm0.1$ & $3.54^{+0.02}_{-0.01}$ & $3.08\pm0.03$\\
       
       & 80102002010 & $17.48\pm0.05$ & $5.97\pm0.02$ &
       $2.93\pm0.01$
       \\
       & 80102002008 & $32.6^{+0.1}_{-0.2}$& $10.98\pm0.03$ &
       $2.97^{+0.01}_{-0.02}$
       \\
       & 80102002006 & $101.9_{-0.2}^{+0.3}$ &
       $35.3\pm0.1$ &
       $2.89\pm0.01$
       \\
       & 80102002004 & $150.1\pm0.2$ & $50.9\pm0.2$ &
       $2.95\pm0.01$\\

       & 80102002002 &      
       $300.9\pm0.3$ & $110.0\pm0.2$ &
       $2.736\pm0.006$\\

\hline
\end{tabular}
}
\end{table*}

\begin{table*} %[h]
\centering
\caption{Best-fit parameters for the broken-linear approximation of the bolometric correction factor $K_{\rm bol}$ as a function of 0.5--10\,keV flux. For linear fits, only $m_1$ and $c_1$ are listed.}
\label{tab:kbol_params}
\begin{tabular}{lcccc}
\hline
\hline
Object & $F_{\rm br}$ & $m_1$ & $c_1$ & $m_2$ \\
       & ($10^{-10}$ erg\,s$^{-1}$\,cm$^{-2}$) & ($10^{-10}$ erg\,s$^{-1}$\,cm$^{-2}$)$^{-1}
       $ &&($10^{-10}$ erg\,s$^{-1}$\,cm$^{-2}$)$^{-1}$\\
\hline
\sw       & $9.9 \pm 1.4$ & $0.09 \pm 0.03$ & $1.9 \pm 0.2$ & $-0.00016\pm0.00006$ \\
\src      &               & $0.0030\pm0.0001$ & $1.95\pm0.01$ &  \\
\smc      &               & $-0.1129\pm0.0001$ & $2.87\pm0.01$ &  \\
\srcV     &               & $-0.0023\pm0.0001$& $2.99\pm0.01$&  \\
\hline
\end{tabular}
\end{table*}

To convert the observed light curves from the 0.5--10~keV band to the bolometric 0.5--79~keV range, we derived bolometric correction factors based on spectral fitting of all archival \textit{NuSTAR} data for each source (see Table~\ref{tab:obsids}). 
The bolometric correction $K_{\rm bol}$ is defined as the ratio between the unabsorbed flux in the 0.5--79~keV band and the 0.5--10~keV band. 

All spectra except for \srcV were re-binned to contain at least 30 counts per energy bin using the \texttt{grppha} utility and fitted with chi-squared statistics in \textsc{xspec} v. 12.15.0 \citep{Arnaud1996}. 
In the case of \srcV, Gehrels weighting \citep{Gehrels1986} was applied, following the methodology of \citet{Doroshenko2017}. 
Unless otherwise specified, all errors are reported at the $1\sigma$ confidence level. 
All fluxes are unabsorbed.

The \nustar data reduction followed the official pipeline recommendations,\footnote{\url{https://heasarc.gsfc.nasa.gov/docs/nustar/analysis/nustar_swguide.pdf}} using \textsc{heasoft} v. 6.35.1 and calibration database \textsc{caldb} v. 20250415. Spectra and light curves were extracted using the \texttt{nuproducts} task from the \texttt{nustardas} package.

\begin{figure}
\centering
\includegraphics[width=0.95\linewidth]{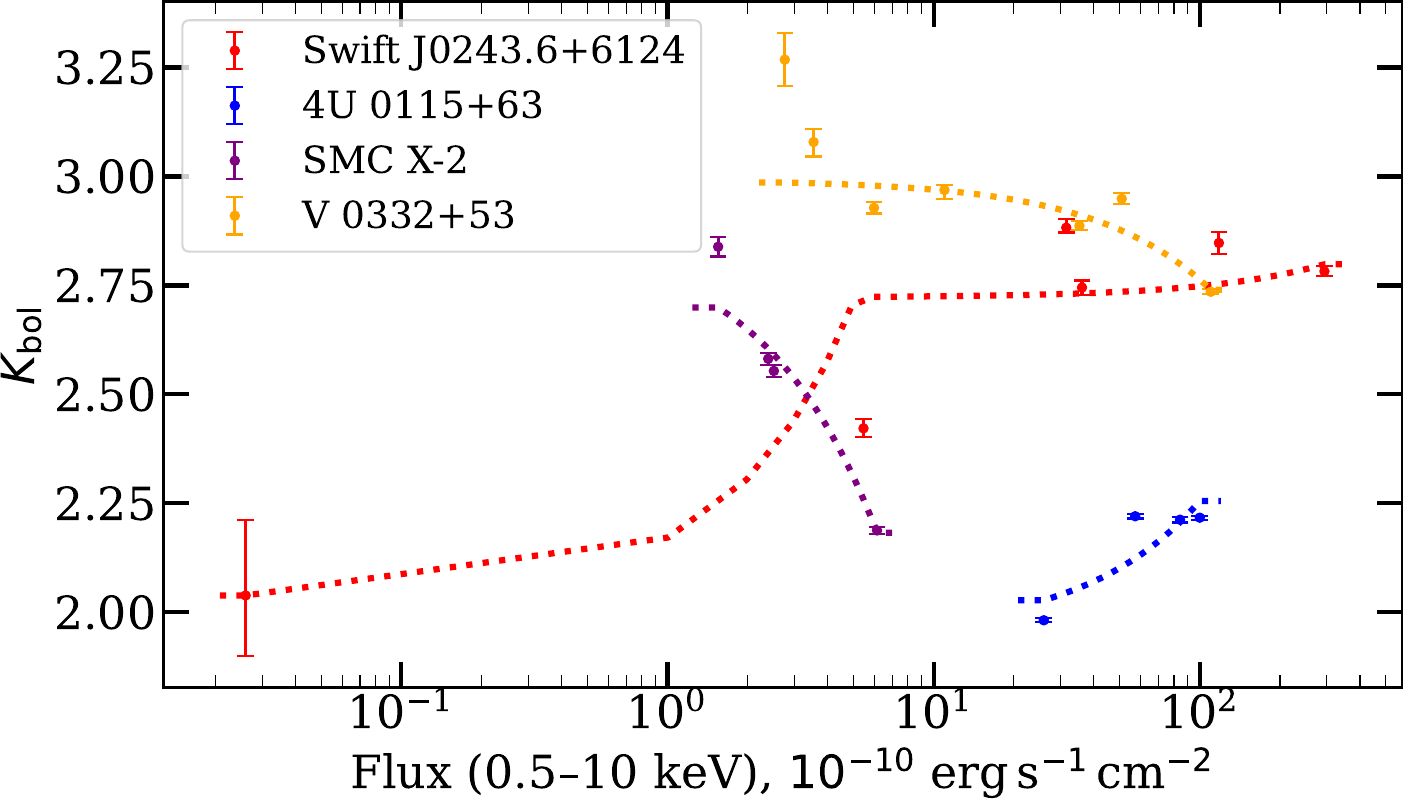}
\caption{Bolometric correction factor $K_{\rm bol}$ as a function of flux in the 0.5--10\,keV band for different XRPs. The piecewise or linear fits for each source are shown with dotted lines.}
\label{fig:bolometric}
\end{figure}

In all cases, the dependence of the bolometric correction factor $K_{\rm bol}$ on the 0.5--10~keV flux ($F_{0.5-10}$) was described using the following broken-linear functional form with constant extrapolation outside the range of observed fluxes:
\begin{equation}
K_{\rm bol}(F) =
\begin{cases}
K_{\rm bol}(F_{\min}),
& F < F_{\min}, \\[1ex]
m_1 F + c_1,
& F_{\min} \le F \le F_{\rm br}, \\[1ex]
m_2 (F -F_{\rm br}) + m_1 F_{\rm br} + c_1 ,
& F_{\rm br} < F \le F_{\max}, \\[1ex]
K_{\rm bol}(F_{\max}),
& F > F_{\max} .
\end{cases}
\label{eq:broken_linear_with_const}
\end{equation}
This function ensures continuity at the break flux $F_{\rm br}$ and provides a conservative, constant extrapolation outside the fitted range. In sources where no break was apparent, the model effectively reduced to a single-slope linear function similarly extrapolated by a constant to low and high fluxes. The parameters of $K_{\rm bol}$ dependence on flux with approximation \eqref{eq:broken_linear_with_const} were determined by least‐squares fitting using the \texttt{curve\_fit} function from the \texttt{scipy.optimize} Python3 library.
The resulting bolometric correction approximations are presented in Fig.~\ref{fig:bolometric}, and the corresponding  parameters are listed in Table~\ref{tab:kbol_params}.

For \sw, the bolometric correction factor was computed using the spectral continuum model $\texttt{cutoffpl} + \texttt{bbodyrad} + \texttt{bbodyrad}$ \citep{Tao2019}, with the number of \texttt{bbodyrad} components varying between observations. 
Two high-flux observations (ObsIDs 90302319004 and 90302319006) exhibited a significant drop in $K_{\rm bol}$, likely due to a transition into the super-Eddington accretion regime \citep{Tao2019}. 
These points were excluded from the fit as they also fall outside the considered flux range.

For \src, we adopted a broadband spectral model that accounts for multiple cyclotron absorption features:
$\mathtt{powerlaw}\times\mathtt{highecut}\times\mathtt{gabs}
\times \mathtt{gabs} \times \mathtt{gabs} \times \mathtt{gabs} \times \mathtt{gabs}$,
where the first \texttt{gabs} component smooths the high-energy cutoff \citep{Coburn2002}, and the remaining four \texttt{gabs} describe the fundamental cyclotron line at $\sim$11~keV and its harmonics at $2E_{\rm cyc}$, $3E_{\rm cyc}$, and $4E_{\rm cyc}$. 

The spectra of \smc were modelled following \citet{Jaisawal2023} using the continuum model
$
\mathtt{cutoffpl}\times\mathtt{gabs}
+\mathtt{bbodyrad},
$ 
where the \texttt{gabs} component accounts for the CRSF at $\sim$27~keV.

For \srcV, the spectral fitting was performed with the model
$
\mathtt{comptt} \times \mathtt{mult\_gabs} \times \mathtt{mult\_gabs},
$
where \texttt{comptt} describes thermal Comptonisation and the two \texttt{mult\_gabs} components represent the CRSF at $\sim$28~keV and its harmonic. 
The \texttt{mult\_gabs} component is defined as
$
G(E) = 1 - D_{\rm cycl} \exp\left[ -\ln 2 \left( \frac{E - E_{\rm cycl}}{\sigma_{\rm cycl}} \right)^2 \right],
$
where $D_{\rm cycl}$ is the line depth, $E_{\rm cycl}$ is the centroid energy, and $\sigma_{\rm cycl}$ is the line width \citep[see][]{Doroshenko2017}.

\newpage 

\section{Details of the  viscous disc models}
\label{sec:freddi_details}

\citet{Tuchman+1990} and \citet{Dubus+1999} showed that a critical level of self-irradiation flux, $Q_{\rm irr} \equiv \sigma_{\rm SB} \,  T_{\rm irr}^4$, is required to maintain a disc ring in the hot state. 
Using a more accurate approach, \citet{Tavleev+2023} calculated the vertical structure of an irradiated disc and found that the critical temperature depends on the ratio of the incident to intrinsic viscous flux (see their Figure 9). 
These authors showed that the irradiation-controlled regime terminates when $T_{\rm irr} \approx 7000$~K. 
The numerical approximation for the found dependence is $T_{\mathrm{irr,crit}} \approx (9040 - 2216 \, Q_\mathrm{vis}/Q_\mathrm{irr} )$~K. 
This approximation  for the critical temperature in the irradiation-controlled regime (when $Q_\mathrm{vis} < Q_\mathrm{irr}$) at the outer hot-zone radius is adopted in the current implementation of \freddi and is used to find the radius $R_{\rm front}$. 
It naturally allows for a smooth transition between the irradiation-controlled regime and the cooling-front propagation, since the critical temperature $T_{\rm eff,min}$ without irradiation is also approximately 7000~K \citep{Lasota+2008}.

In addition to the updated condition on the critical irradiation temperature, \freddi includes a modification of the boundary condition at the hot-zone radius. 
In the DIM framework, a decretion of matter is present from the hot to cold zone \citep{Ludwig-Meyer1998,Menou+1999}. 
The magnitude of this decretion rate is actually somewhat uncertain as it cannot be estimated from observations and depends on the variation of $\alpha$ in the H-recombination zone.  
For the DIM framework,  \citet{Menou+1999} showed that the hot zone evolves in a self-similar fashion, and the mass decretion rate at $R_\mathrm{hot}$ can be expressed as the accretion rate at the inner disc edge multiplied by a constant factor, which we call the `front-decretion' factor. 
We have implemented this boundary condition in the current modification of \freddi, with the front-decretion factor $-1.8$.
This value ensures that the results of the \freddi calculation are consistent with those of the DIM code (Jean-Marie~Hameury, private communication). 

Self-irradiation is treated as described in Sect.\,3.1 of \citet{Avakyan+2024}. 
The self-irradiation parameter defined by  Eq.~\eqref{eq:cirr} is implemented in the form
\begin{equation}
C_{\rm irr}  =  \tilde{C}_{\rm irr}\, \frac{z/r}{0.05}  \, .
\end{equation}
Given the typical value of the relative thickness (see Eq.~\eqref{eq:zr}), $C_{\rm irr}$ and $\tilde{C}_{\rm irr}$ are generally close to each other.
For the present calculations, we assume that the angular distribution of the central radiation is isotropic.

The fits presented in Fig.~\ref{fig:models_freddi} were performed by locating the global maximum using the Levenberg–Marquardt method. The main source of uncertainty in the resulting parameters stems from the choice of parametric form of $C_{\rm irr}$, as well as from the specific conditions at the cold-front radius, as described above, and from the uncertainty in the distances to the sources. Moreover, as mentioned in Sect.~\ref{sec:dim}, there is a degeneracy between $\alpha$  and $C_{\rm irr}$.  Therefore, reporting uncertainties of parameters would be misleading and constitute an over-interpretation of the data. 

\end{appendix}
\end{document}